\documentclass[aps,twocolumn,showpacs,prb,superscriptaddress]{revtex4-2}
\usepackage{mathtools}
\usepackage{times}
\usepackage{bm}
\usepackage{dsfont,amsthm,amsbsy}
\usepackage{verbatim}
\usepackage{amssymb}
\usepackage{amsmath}
\usepackage{bbm}
\usepackage{graphicx}
\usepackage{epstopdf}
\usepackage{subfigure}
\usepackage{natbib}
\usepackage{epsfig}
\usepackage{amsfonts}
\usepackage{mathrsfs}
\usepackage{sidecap}
\usepackage{lipsum}
\usepackage[toc,page,title,titletoc,header]{appendix}
\usepackage[colorlinks,linkcolor=blue,citecolor=blue,anchorcolor=blue,urlcolor=blue]{hyperref}
\usepackage{hyperref}
\usepackage{resizegather}
\usepackage{tikz}
\usepackage{float}
\usepackage{mathbbol}
\usepackage[normalem]{ulem}
\usepackage{cancel}
\usepackage{upgreek}

\newcommand{\bl}{\begin{aligned}}
\newcommand{\el}{\end{aligned}}

\def\be{\begin{equation}}
\def\ee{\end{equation}}

\def\bi{\begin{itemize}}
\def\ei{\end{itemize}}
\def\bn{\begin{enumerate}}
\def\en{\end{enumerate}}
\def\bea{\begin{eqnarray}}
\def\eea{\end{eqnarray}}
\def\no{\nonumber}
\def\ba{\begin{array}}
\def\ea{\end{array}}
\def\bd{\begin{displaymath}}
\def\ed{\end{displaymath}}

\def\tr{{\rm tr}}

\begin{document}
\title{Noise-Affected Dynamical Quantum Phase Transitions}


\author{R. Jafari}
\email[]{raadmehr.jafari@gmail.com}
\affiliation{Department of Physics, University of Gothenburg, SE 412 96 Gothenburg, Sweden}

\author{Alireza Akbari}
\affiliation{Beijing Institute of Mathematical Sciences and Applications (BIMSA), Huairou District, Beijing 101408, China}
\author{Mehdi Biderang}
\affiliation{Department of Physics, University of Toronto, 60 St. George Street, Toronto, Ontario, M5S 1A7, Canada}
\affiliation{DelQuanTech Inc., 500 Doris Ave., Toronto, Ontario, M2N 0C1, Canada}

\author{Jesko Sirker}
\affiliation{Department of Physics and Astronomy, University of Manitoba, Winnipeg R3T 2N2, Canada}
\affiliation{Manitoba Quantum Institute, University of Manitoba, Winnipeg R3T 2N2, Canada}


\begin{abstract}
We investigate the effects of uncorrelated noise on dynamical quantum phase transitions (DQPTs) in fermionic two-band models following a quantum ramp across critical points. We consider a generalized Loschmidt echo for the noise-averaged density matrix $\bar\rho$, which is a mixed state in general, as well as the pure state Loschmidt echo calculated for each noise realization with the average performed over the corresponding return rates. $\bar\rho$ can be obtained from a master equation and we show that for two-band models noise destroys its coherences which typically drives $\bar\rho$ towards the completely mixed state which is an attractive fixed point. DQPTs are thus always smoothed out for finite noise. For single noise realizations, on the other hand, we find that DQPTs under certain conditions are always present irrespective of the noise level. This leads to remarkable stable though slightly broadened DQPT-like features in the averaged return rate. We illustrate our results for the XY model by considering a noisy ramp as well as noise in the energy levels of the final Hamiltonian.
\end{abstract}

\maketitle

\section{Introduction}
In the last decade, there has been renewed interest in the non-equilibrium dynamics of quantum systems largely driven by progress in various experimental platforms ranging from ultracold atomic gases and trapped ions to atom-cavity systems \cite{kasprzak2006,Kessler2019,Klinder2015,Fitzpatrick2017,Smale2019,Zhang2017,Nicklas2015,Prufer2018,Erne2018}. These systems allow for a precise control at the atomic level making it possible to study various types of non-equilibrium dynamics. 

A particular focus of experimental and theoretical studies have been quantum quenches and quantum ramps where an external parameter is either changed suddenly or over some time interval. In the ensuing dynamics one can study, for example, the dynamics of an order parameter or one can consider how much the time evolved state deviates from the initial state which is captured by the Loschmidt echo. There are two different notions of dynamical quantum phase transitions (DQPTs) in this scenario: One can consider the vanishing of an order parameter or a correlation function as defining a DQPT \cite{SchutzholdUhlmann,MarinoEckstein,Heyl2018,MarinoEckstein,Tian2020} or one can define DQPTs as zeroes in the Loschmidt echo \cite{Heyl2013,Andraschko2014,Karrasch2013,Heyl2015,Heyl2018,Karrasch2017,Sedlmayr2018,SedlmayrJaeger,Sedlmayr2023,MaslowskiCheraghi}. The former is sometimes referred two as DQPTs of type I and the latter as DQPTs of type II. It is important to note that the two are distinct in general. In this paper we will concentrate entirely on DQPTs of type II which have been observed directly on different experimental platforms \cite{Jurcevic2017,Guo2019,Wang2019,Nie2020}.

A largely open question is how noise, which can be viewed as an efficient way of describing the evolution of systems interacting
with environments or external driving fields and which is ubiquitous in nature, will affect DQPTs. Although DQPTs in a system with a stochastically driven field have been studied before \cite{Jafari2024}, a detailed analysis of the noise average has not been performed yet. In this paper we will address this question for fermionic two-band models and illustrate our results using the XY model as an example. An important point we stress in our study is that the noise average can be performed at different stages of the measurement, leading to qualitatively different results. On the one hand, we can time evolve a pure state in the presence of noise and then calculate the noise-averaged state $\bar\rho$ which will be a mixed state. In this case a generalized Loschmidt echo for mixed states has to be used to provide a measure of the distance between the time evolved mixed state and the initial state. The advantage of this approach is that a Lindblad-type master equation can be derived for $\bar\rho$ which we can either solve analytically in certain cases or use to obtain a numerical solution efficiently. However, experimentally it is challenging to completely characterize a state e.g.~by quantum tomography. From this perspective, it seems more feasible to obtain the pure-state Loschmidt echo and the corresponding return rate for individual noise realizations and to perform the noise average at the end. In particular, if the initial state $|\Psi(0)\rangle$ is a simple product state then obtaining the pure state Loschmidt echo $\langle\Psi(0)|\Psi(t)\rangle$ amounts to a simple projection. We find that in a quench across a critical point in fermionic two-band models, DQPTs do always occur for any noise level under certain conditions. The broadening of DQPTs in the noise-averaged return rate is then the consequence of the critical times at which DQPTs do occur being different for each noise realization. If the noise levels are small to moderate, the shifts in the critical times are often relatively small between different realizations, leading to surprisingly sharp DQPT-like features even in the presence of noise. 

Our paper is organized as follows: In Sec.~\ref{Sec_Loschmidt}, we review the Loschmidt echoes and corresponding return rates both for the cases of pure and of mixed states. For the noise-averaged density matrix we derive a Lindblad-type master equation. We also introduce the ramp and measuring protocol which we will analyze in the following. In Sec.~\ref{Sec_two-bands}, we present a general analysis of the two return rates for a general fermionic two-band model. In Sec.~\ref{Sec_XY}, we illustrate our general results buy using the XY model as an example. We consider both quenches across a single and across two critical points. Furthermore, we consider two noise scenarios: In the first, noise is present during the ramp while in the second the ramp is noiseless but there is noise in the energy levels of the final Hamiltonian. In Sec.~\ref{Sec_Concl}, we summarize our main results and conclude.

\section{Ramp protocol and Loschmidt echoes}
\label{Sec_Loschmidt}
We start by describing the measuring protocol we have in mind and review the definitions of the Loschmidt echo for pure and for mixed states. While these definitions are general, we describe the ramp and measuring protocol first to be able to directly introduce the notation which will be used in the rest of the paper. 
\begin{figure}[htp!]
    \includegraphics[width=0.99\columnwidth]{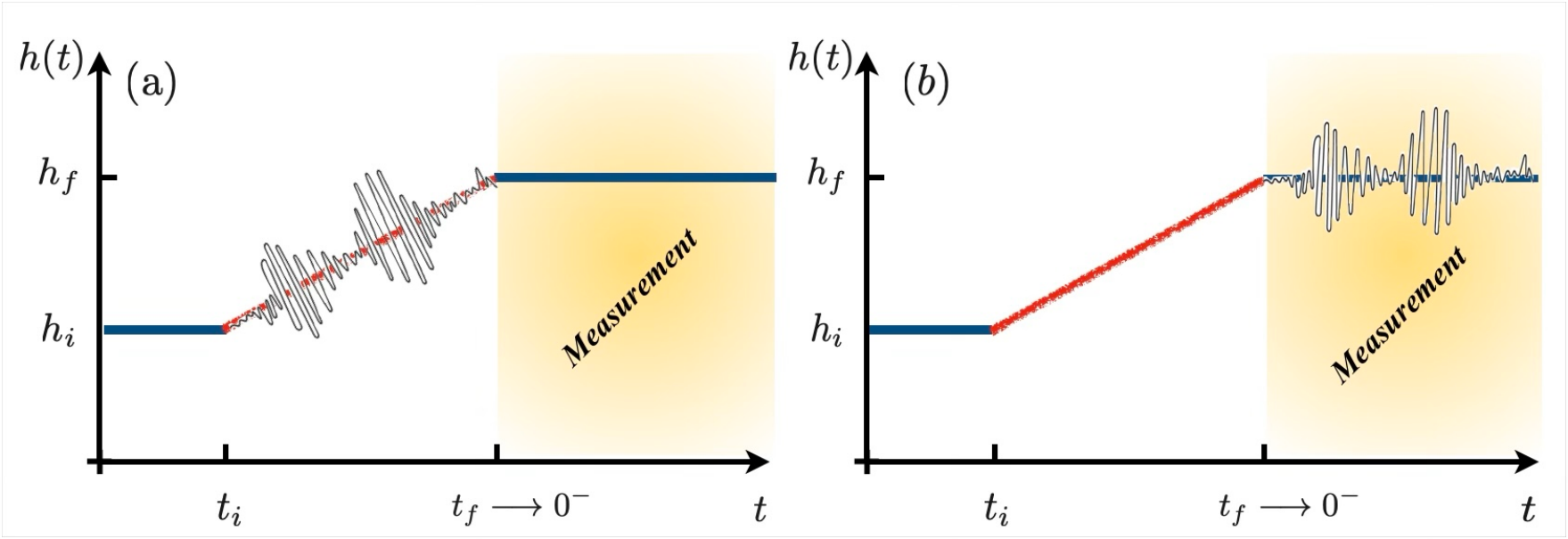}
    \caption{Two types of noise are being considered: (a) A ramp where the ramp field $h$ is noisy, and (b) a noiseless ramp where energy fluctuations are, however, present in the final Hamiltonian. In both cases the Loschmidt echo is measured for times $t>0$.}
    \label{Fig_schematics}
\end{figure}
This protocol is shown in Fig.~\ref{Fig_schematics} and consists of the following steps: (i) At the initial time $t_i$, the system is prepared in the ground state $|\Psi^i(t_i)\rangle$ of the Hamiltonian $H(h_i)$ where $h$ is an external control parameter such as a magnetic field. (ii) During the time interval $[t_i,t_f)$ the field $h$ is increased in a linear ramp. This ramp can either be noisy (Fig.~\ref{Fig_schematics}(a)) or noiseless (Fig.~\ref{Fig_schematics}(b)). The state of the system after the ramp, $|\Psi(t_f\to0^-)\rangle$, is then time evolved further with the final Hamiltonian $H(h_f)$ and the Loschmidt echo is measured as a function of time $t>0$. We consider both the case where $H(h_f)$ is noiseless and the case where $H(h_f)$ contains noise in the final energy levels. 

For each noise realization $n$, the initial state at time $t_i$ is always the same, $|\Psi^i_n(t_i)\rangle\equiv|\Psi^i\rangle$. The time evolution during the ramp is then described by a Schrodinger equation $|\dot\Psi^i_n(t)\rangle=-iH^i_n(t)|\Psi^i_n(t)\rangle$ (we set $\hbar=1$) with a time-dependent Hamiltonian $H^i_n(t)$ which can contain noise, $H^i_n(t)=H^i(h(t))+R^i_n(t)$, where $H^i(h(t))$ describes the noiseless ramp and $R^i_n(t)$ is the $n$-th noise realization. At the end of the ramp at time $t_f$, the system is in the state $|\Psi^f_n(0)\rangle\equiv |\Psi^i_n(t\to 0^-)\rangle$. For times $t>0$, this state is further time-evolved and fulfills the Schrodinger equation $|\dot\Psi^f_n(t)\rangle=-i H^f_n(t)|\Psi^f_n(t)\rangle$ with $H^f_n(t)=H^f(h_f)+R_n^f(t)$. We then measure the Loschmidt echo between the state $|\Psi_n^f(t=0)\rangle$ and the state $|\Psi_n^f(t>0)\rangle$. For a single noise realization $n$, both are pure states and we can define the Loschmidt echo as  
\begin{equation}
\label{L_pure}
L^2_n[\rho_n^f(0),\rho_n^f(t)]=|\langle\Psi^f_n(0)|\Psi^f_n(t)\rangle|^2=\tr\left(\rho^f_n(0)\rho^f_n(t)\right) \, .
\end{equation}

While we have written the Loschmidt echo in Eq.~\eqref{L_pure} in terms of density matrices, this form of the Loschmidt echo only induces a proper metric if the density matrices are pure. A more general definition which also holds for mixed-state density matrices is given by \cite{Sedlmayr2018}
\begin{equation}
    \label{L_mixed}  \mathcal{L}_n[\rho_n^f(0),\rho_n^f(t)]=\tr\sqrt{\sqrt{\rho_n^f(0)}\rho_n^f(t)\sqrt{\rho_n^f(0)}} \; .
\end{equation}
Note that Eq.~\eqref{L_mixed} reduces to Eq.~\eqref{L_pure} if the density matrices are pure. The corresponding return rate is then defined by
\begin{equation}
    \label{return_rate}
    g_n[\rho_n^f(0),\rho_n^f(t)]=-\lim_{N\to\infty}\frac{1}{N}\ln|\mathcal{L}_n[\rho_n^f(0),\rho_n^f(t)]|^2
\end{equation}
and takes care of the fact that the overlap between the density matrices decreases exponentially with system size $N$ due to the usual orthogonality catastrophy.

We can now define noise averages in two different ways: We can either calculate the return rate for each noise realization and only at the very end take the average or we can calculate the noise-averaged density matrices first and then calculate the Loschmidt echo for these mixed-state density matrices. This leads to the two definitions:
\begin{eqnarray}
    \label{noise_average}
    \bar{g}[\rho]&\equiv&\bar{g}[\{\rho^f(0),\rho^f(t)\}] = \lim_{M\to\infty}\frac{1}{M}\sum_n g_n[\rho_n^f(0),\rho_n^f(t)] \nonumber \\  g[\bar{\rho}]&\equiv&g[\bar{\rho}^f(0),\bar{\rho}^f(t)],\;\mbox{with}\;\; \bar{\rho}^f(t)\!=\!\lim_{M\to\infty}\!\frac{1}{M}\sum_n \rho_n^f(t)  
\end{eqnarray}
The first definition $\bar g[\rho]$ appears more accessible from an experimental point of view because it only requires the preparation of the initial state $|\Psi^{i}\rangle$, which is typically a simple product state, and the measurement of the pure-state Loschmidt echo \eqref{L_pure}. We note, however, that the latter might also be non-trivial because $\rho_n^f(0)$, while being pure, is in general no longer a product state. The second definition $g[\bar\rho]$, on the other hand, does require the use of the generalized Loschmidt echo \eqref{L_mixed} but has the theoretical advantage that a master equation for the noise-averaged density matrix $\bar\rho^f(t)$ can be derived which can either be solved analytically or otherwise allows at least for a very efficient numerical solution.

\subsection{Master equation for the noise-averaged density matrix}
Let us first discuss how a master equation for the noise-averaged density matrix can be obtained. We begin by considering a general time-dependent Hamiltonian,
%
\begin{equation} 
\label{NH}
H(t)=H_{0}(t)+r(t)H_{1},
\end{equation}
%
where $H_{0}(t)$ is noise-free and $r(t)$ a real function for a given realization of the noise. This expression for $H(t)$ well captures linear corrections from a weak stochastic variation. As noted in Ref.~\cite{Kiely2021sm}, the resulting formalism can readily be adapted to apply also beyond the linear regime.

Here we consider Gaussian noise $r(t)$ with mean $\langle r({t})\rangle=0$. The prototype is Ornstein-Uhlenbeck (colored) noise \cite{CoxMiller1965sm}, which
is a stochastic process with correlation function
%
\begin{equation}
\label{CNC}
\langle r({t})r({t}')\rangle=\frac{\xi^{2}}{2\tau_n}e^{-|{t}-{t}'|/\tau_n}
\end{equation}
%
where $\xi$ is the amplitude of the noise, $\tau_n$ is the noise correlation time, and the limit $\tau_n \rightarrow 0$ defines Gaussian white noise with the correlation function 
%
\begin{equation}
\label{WNC}
\langle r({t})r({t}')\rangle = \xi^2\delta(t-t')
\end{equation}
%
which will be our focus here. The colored noise master equation for the averaged density matrix $\bar\rho(t)$ of $H(t)$ 
is given by
%
\begin{eqnarray}
\label{CNmaster}
\dot{\bar\rho}(t)&=&-i[H_{0}(t),\bar\rho(t)] \\
&-&\frac{\xi^{2}}{2\tau_n}\Big[H_{1},\int_{t_i}^{t}e^{-(t-s)/\tau_{n}}[H_{1},\bar\rho(s)]ds\Big]. \nonumber
\end{eqnarray}
%
which reduces to the white noise master equation in the limit $\tau_n \rightarrow 0$ \cite{Asadian2025},
%
\begin{equation}
\label{Master}
\dot{\bar\rho}(t)=-i[H_{0}(t),\rho(t)]-\frac{\xi^{2}}{2}\Big[H_{1},\Big[H_{1},\bar\rho(t)\Big]\Big].
\end{equation}
%
We note that this equation has the form of a standard Lindblad master equation with dissipator
\begin{eqnarray}
    \label{dissipator}
    D[\bar\rho] &=& -\frac{\xi^2}{2}\Big[H_{1},\Big[H_{1},\bar\rho\Big]\Big] \nonumber \\
    &=& \xi^2 \left(H_1\bar\rho H_1^\dagger-\frac{1}{2}H_1^\dagger H_1\bar\rho -\frac{1}{2}\bar\rho H_1^\dagger H_1 \right)
\end{eqnarray}
with $H_1^\dagger =H_1$ Hermitian in this case. In the following, we will analyze the white-noise master equation \eqref{Master} for different scenarios and compare the results obtained for the return rate $g[\bar\rho]$ with results for the averaged pure-state return rate $\bar g[\rho]$.

\section{Two-band models}
\label{Sec_two-bands}
To further simplify the discussion, we focus on an integrable, one-dimensional systems that can be 
reduced to a two-level fermionic Hamiltonian $H_k(h)$ for each momentum mode $k$, with a tunable parameter $h$.
Such systems can serve as a paradigm for exploring quantum and topological phase transitions in and out of equilibrium,
and represent several generic spin chains and fermionic models for suitably chosen parameters. We focus first on a single measurement of the return rate either for a time evolution without noise or for a single noise realization during the ramp (scenario (a) in Fig.~\ref{Fig_schematics}).

At the initial time $t_{i}$, the system is prepared in the ground state $|\Psi^i\rangle=\prod_k| \Psi^{i}_{k}\rangle $ 
of the initial Hamiltonian $H^i=\sum_k H_{k}(h_{i})$. 
In the first stage of the protocol a linear ramp in the parameter $h$ from an initial value $h_i$ at $t_i$ to a final value $h_f$ at time $t_f \to 0^{-}$ takes place. An adiabatic evolution condition breaks down when crossing a critical (gap closing) point at a finite speed $v$. Therefore the final state after the ramp, $|\Psi^f\rangle=\prod_k |\Psi_k^f\rangle\equiv|\Psi(t\to 0^{-})\rangle$, is not the ground state of the post-ramp Hamiltonian $H^{f} =\sum_k H_{k}(h_f)$.
Instead, it is in general given by a linear combination $| \Psi_{k}^{f} \rangle = v_{k} |\alpha_{k}^{f}\rangle +u_{k} | \beta_{k}^{f}\rangle $ with $|v_{k}|^{2}+|u_{k}|^{2}=1$ where $|\alpha_{k}^{f}\rangle$ and $|\beta_{k}^{f} \rangle$ are the ground and the excited states of the two-level post-ramp Hamiltonian $H_{k}^{f}$, respectively, with the corresponding energy eigenvalues assumed to be $\pm\epsilon^{f}_{k}$. The probability of a non-adiabatic transition, which results 
in the system being in the excited state at the end of the ramp, is then given by $p_{k}=|u_{k}|^{2}=|\langle\beta_{k}^{f}|\Psi_{k}^{f}\rangle |^{2}$. 
In the second stage of the protocol, the Loschmidt amplitude $\mathcal{L}(t)=\prod_k \mathcal{L}_k(t)$ for $t>0$ is considered which, for a two-band model, is given by \cite{Zamani2024,Sharma2016}
\begin{equation}
{\cal{L}}_{k}(t) = \langle \psi^{f}_{k}|\text{e}^{-iH^{f}_{k}t} |\psi^{f}_{k} \rangle 
= |v_{k}|^{2}e^{i\epsilon^{f}_{k}t}+  |u_{k}|^{2}e^{-i\epsilon^{f}_{k}t} \, .
\label{eq1}
\end{equation}
%
Since we are here only considering pure states, we use the notation $\mathcal{L}(t)\equiv \mathcal{L}[\rho^f(0),\rho^f(t)]$, i.e., we omit to explicitly include the dependence on the pure state density matrices. The return rate then reads $g(t)=\lim_{N\to\infty}\frac{1}{N}\sum_k g_k(t)$ with $g_{k}(t)  = - \ln |{\cal{L}}_{k}(t)|^{2}$ where $N$ is the size of the system \cite{Heyl2013,Heyl2018}. Converting the sum to an integral in the thermodynamic limit one finds \cite{Sharma2016}
%
\begin{equation}
\label{eq:DFE}
g(t)=-\frac{1}{\pi} \int_{0}^{\pi}\ln\left(1-4p_{k}(1-p_{k})\sin^{2}(\epsilon_{k}^{f}t)\right) dk.
\end{equation}	    	
%
The Loschmidt amplitude $\mathcal{L}(t)$ vanishes if any of the factors $\mathcal{L}_k(t)$ vanishes which happens at
\begin{equation}
\label{Fisher}
z_n(k)=it_n(k)=\frac{i\pi}{\epsilon_k^f}\left(n+\frac{1}{2}\right)+\frac{1}{2\epsilon_k^f}\ln\left(\frac{p_k}{1-p_k}\right)
\end{equation}
with $n=0,1,2,\cdots$. These so-called Fisher zeroes form curves in the complex plane in the one-dimensional case considered here. They cross the imaginary axis---corresponding to real critical times---if there are critical momenta $k^\ast$ such that $p_{k^{\ast}}=1/2$. In this case the critical times are 
%
\bea \label{eq:criticaltimes}
t_n^{\ast} = t^{\ast} \left(n+\frac{1}{2}\right), \quad t^{\ast}=\frac{\pi} {\varepsilon^f_{k^{\ast}}};  \quad n=0,1,2,\cdots.
\eea
%
Note that $\mathcal{L}_{k^\ast}(t^\ast_n)=0$ also corresponds to the times and momenta where the argument of the logarithm in Eq.~\eqref{eq:DFE} vanishes. This leads to non-analyticities in $g(t)$ or its time derivatives at $t=t^\ast$ which are called DQPTs. We remind the reader again that the term DQPT is also used when the behavior of order parameters or correlation functions under a quench or ramp across a quantum critical point is studied \cite{SchutzholdUhlmann,MarinoEckstein,Heyl2018}. In many cases there is, however, no direct relation to the DQPTs defined above and studied here. 

The Fisher zeroes form curves $z_n(k)$ in the complex plane as a function of momentum $k$, labeled by the integer $n$, which typically cross the imaginary axis at some angle $\delta$. If the density of Fisher zeroes near such a critical time $t^\ast_n$ where the imaginary axis is crossed is constant, then this causes a jump in the first derivative of the return rate, $\dot g(t)\sim \pm\cos(\delta)$. However, more exotic scenarios where $\dot g(t)$ diverges for $t\to t^\ast_n$ are in principle possible as well if the density of Fisher zeroes diverges at $t_n^*$, see Ref.~\cite{MaslowskiCheraghi} for a detailed analysis of the different possible scenarios. The most common case of a jump in $\dot g(t)$ has also been found for a noiseless ramp in a spin chain for all values of the ramp velocity~\cite{Zamani2024}.

To set the stage for investigating the effects of noise, we consider the fermionic two-band model described by the Hamiltonian
%
\begin{equation}
\label{eq:Nambu}
{\cal H}^0_k(t) = \sum_{k} C^{\dagger}_k {\cal H}_k^{0}(t) C_k;
\;
{\cal H}^{0}_k(t)=
\begin{pmatrix}
h^{z}_k(t) & \gamma_k \\
\gamma_k & -h^{z}_k(t)
\end{pmatrix}
\end{equation}
%
using the Nambu spinor $C_k^{\dagger} = (c_k^{\dagger} \ c_{-k})$. Here $\pm h^z_k(t)$ are the bare energy bands of the two levels which will change during the ramp. $\gamma_k$ describes a momentum dependent but time independent mixing between the two levels. More specifically, we can write $h^z_k(t)=h_0(t)+f_1(k)$ and $\gamma_k=\gamma f_2(k)$ with $\gamma\neq 0$. Here, $h_0(t_i)=h_i$, $h_0(0)=h_f$ is the time-dependent field and $f_{1,2}(k)$ are momentum-dependent functions which are determined by the microscopic details of the model under investigation. The ramp $h_i\to h_f$ will cross a critical, gap-closing point at which an adiabatic approximation breaks down at $h_0(t)=-f_1(\tilde k)$ if $f_2(\tilde k)=0$. Note that there can be multiple momenta $\tilde k$ for which this condition is fulfilled. 
Linear ramps from $h_i\to -\infty$ to $h_f\to +\infty$ or vice versa are particularly easy to analyze because we can define $h^z_k(t)=h_f+vt+f_1(k)\equiv h_{f,k}+vt$ with $h_{f,k}\to\infty$ and the details of the bare dispersion $f_{1}(k)$ thus become irrelevant. 

Let us first review the case of a noiseless ramp in this limit, starting from the ground state of the Hamiltonian $\mathcal{H}_k^0(t_i)$.
The transition rate can then be calculated by the Landau-Zener formula $p_k=\exp(-\pi\gamma_k^2/v)$ \cite{Landau,Zener}. DQPTs occur if $p_{k}\leq 1/2$ for some $k$ \cite{Sharma2016,Zamani2024}. From the Landau-Zener formula we see that this condition can only be satisfied for ramp velocities $v\leq v_c=\pi \max(\gamma_k^{2})/\ln(2)$. For higher velocities and sudden quenches no DQPTs will appear \cite{Sharma2016,Zamani2024,Jafari2024}. If the ramp is very slow, a particularly simple picture emerges. While adiabaticity is always broken at the critical points, the ramp will then transfer most of the weight from the old ground state $|\Psi_k(t_i)\rangle$ to the new ground state of the system if $\gamma_k\neq 0$. The exact solution, see App.~\ref{App_A}, for the probability $p_k$ to be in the excited state of the final Hamiltonian can then be approximated by $p_k\approx 1/2-1/2\tanh(vt+h_f)$. At time $t_0=-h_f/v$ where the field $h_0(t_0)=0$ the system is then approximately described by the pure-state density matrix $\rho(t_0)$ with diagonal elements $1/2$ and off-diagonal elements (coherences) $-1/2$. As we will argue in the following, noise in the ramp field $h_0(t)$ can destroy the coherences and make the system flow to the completely mixed state $\bar\rho_k=\mbox{diag}(1/2,1/2)$ if we base our measurements on the noise-averaged density matrix.

\subsection{Noise-averaged density matrix}
For a two-level system, the total density matrix is a direct product of the density matrices $\rho_k$ for each $k$-mode. The master equation \eqref{Master} thus separates
%
\begin{equation}
\bl
\label{eq:master}
\frac{d}{dt}\bar{\rho}_{k}(t)=-i[H_{k}^0(t),\bar{\rho}_{k}(t)]-\frac{\xi^2}{2}[H_1,[H_1,\bar{\rho}_{k}(t)]].
\el
\end{equation}
%
We want to study, in particular, the case where $H_1=\sigma^z$ relevant for the case where we have noise in the external driving field $h$. We make the following general observations: (1) $\bar\rho_k(t_i)=|\Psi_k(t_i)\rangle\langle\Psi_k(t_i)|$ always commutes with $H_1$ and also does commute with $H_{k}^0(t)$ if $\gamma_k=0$. Thus, modes with $\gamma_k=0$ are stationary. (2) The completely mixed state $\bar\rho_k=\mbox{diag}(1/2,1/2)$ is a fixed point of Eq.~\eqref{eq:master}. The question then is if this fixed point is attractive. We first note that there is a time interval of length $\sim 1/v$ around $t_0=-h_f/v$ where $|h_0(t)|\ll 1$. In this time interval, where the dynamics mainly takes place, we can drop $h_0(t)$ from the Hamiltonian. To check whether the fixed point is attractive, we now write the general density matrix $\bar\rho_k$ at time $-h_f/v$ as having diagonals $1/2\pm r(t)$ and coherences (off-diagonal elements) $\varepsilon(t)$ and $\varepsilon^\ast(t)$, respectively. Eq.~\eqref{eq:master} then leads to the coupled differential equations $\dot r =-2\gamma \text{Im}(\varepsilon)$ and $\dot\varepsilon = 2i\gamma r-2\varepsilon\xi^2$. One sees that $r(t\to\infty)=0$ and $\varepsilon(t\to\infty)=0$, i.e., the fixed point is always attractive for $\xi\neq 0$. However, the dynamics becomes frozen for times $|t-t_0|\gg 1/v$ because then $|h_0(t)|$ becomes large. The question therefore is whether the completely mixed fixed point is reached before this happens. For $\xi\lesssim \gamma$ the approximate solution of the differential equations above is $r/\varepsilon(t)\approx r/\varepsilon(0)\exp(2i\gamma t)\exp(-\xi^2 t)$, respectively, and the fixed point is reached if $\xi^2/v\gg 1$. Conversely for $\xi\gg \gamma$, $r(t)\approx r(0)\exp(-2t \gamma^2/\xi^2 )$  while $\varepsilon(t)\approx \varepsilon(0)\exp(-2\xi^2 t)$ decays much faster. In the latter case, the fixed point is reached if $\gamma^2/(\xi^2 v)\gg 1$. To summarize, the completely mixed state is always reached if $\gamma_k\neq 0$ and the ramp is sufficiently slow thus representing a universal phenomenon in noisy ramps. Since the completely mixed state is a fixed point of the master equation, no further time evolution will occur if $\bar\rho_k(0)=\mbox{diag}(1/2,1/2)$. Consequently, $\mathcal{L}[\bar\rho_k(0),\bar\rho_k(t)]=1$. If this is the case for almost all $k$, then the return rate becomes $g[\bar\rho](t)\equiv0$. We therefore expect that with increasing noise level $\xi$ the return rate $g[\bar\rho](t)$ at a given time $t$ is monotonically decreasing and becomes zero for large noise.

\subsection{Noise-averaged return rate}
We note first that for each individual noise realization, Eq.~\eqref{eq:DFE} remains valid and the effect of noise is encapsulated entirely in the function $p_k$. For momenta $k$ for which the mixing rate $\gamma_k$ between the two levels is zero, $p_k$ is determined entirely by the eigenstates of the Hamiltonian before and after the ramp and will be given by $p_k=0$ or $p_k=1$. DQPTs occur even in the presence of noise if a momentum $k^*$ exists with $p_{k^*}=1/2$. Such a momentum is guaranteed to exist if $\gamma_k$ vanishes for two $k$-modes such that for one $p_k=0$ while it is given by $p_k=1$ for the other. Thus, noise does in general not prevent DQPTs from occuring. However, the momenta $k^*$ for which $p_{k^*}=1/2$ will be different for each noise realization thus leading, according to Eq.~\eqref{eq:criticaltimes}, to different critical times $t^*$ for each realization. In the noise-averaged return rate $\bar g[\rho]$ this will lead to DQPTs being smoothed out. However, for small noise levels the momenta $k^*$ are expected to shift only slightly, leading to DQPT-like features which are still quite sharply defined. In contrast to the return rate $g[\bar\rho]$ based on the noise-averaged density matrix, the behavior of $\bar g[\rho]$ does not have to be monotonic as a function of noise. In particular, $p_k$ will in general not vanish or be equal to $1$ for all $k$ no matter how strong the noise is, implying that $\bar g[\rho](t)$ will always be non-zero.

\section{XY model}
\label{Sec_XY}
To ilustrate our general results, we consider now the specific Hamiltonian of an $XY$ model
%
\bea\no
{\cal H}^0(t) 
\!=\!
-\frac{1}{2} \sum_{n=1}^{N} \Big[\frac{1
\!+\!
\gamma}{2}\sigma_n^x \sigma_{n+1}^x \!+\!
\frac{1
\!-\!
\gamma}{2} \sigma_n^y \sigma_{n+1}^y
\!-\!
h_0(t) \sigma_n^z \Big],
\eea
%
where $\sigma^{x,y,z}$ are the Pauli matrices. The Hamiltonian ${\cal H}^0(t)$ can be mapped onto a model of spinless fermions with operators $c_n^{(\dagger)}$ using a Jordan-Wigner transformation \cite{LSM1961}. Performing a Fourier transformation, $c_n = (\mbox{e}^{i\pi/4}/\sqrt{N}) \sum_k \mbox{e}^{ikn}c_k$ (the phase factor $\mbox{e}^{i\pi/4}$ has been added for convenience), we obtain the two-band Hamiltonian \eqref{eq:Nambu} with $h^{z}_k(t) = h_0(t)-\cos(k)$ and $\gamma_k = \gamma\sin(k)$. We have $\gamma_k=0$ for $k=0,\pi$ leading to two critical points $h=\pm 1$. These two critical fields correspond to quantum phase transitions from a paramagnetic to a ferromagnetically ordered phase  \cite{Barouch1970}.

\subsection{Noisy ramp across two critical points}
We first consider a ramp across both critical points. The Landau-Zener formula tells us that for $h_i\to -\infty$ and $h_f\to\infty$ there is a critical velocity $v_c=\pi\gamma^2/\ln 2$ such that no DQPTs will occur if $v>v_c$. This picture will also hold more generally if we ramp from a finite field $h_i< -1$ to a finite field $h_f>1$ with some renormalization of the critical velocity $v_c$. In the following, we will consider the influence of noise both for a ramp with $v<v_c$ and for a ramp with $v>v_c$.

\subsubsection{Averaged density matrix}
\begin{figure}[htp!]
    \includegraphics[width=0.99\columnwidth]{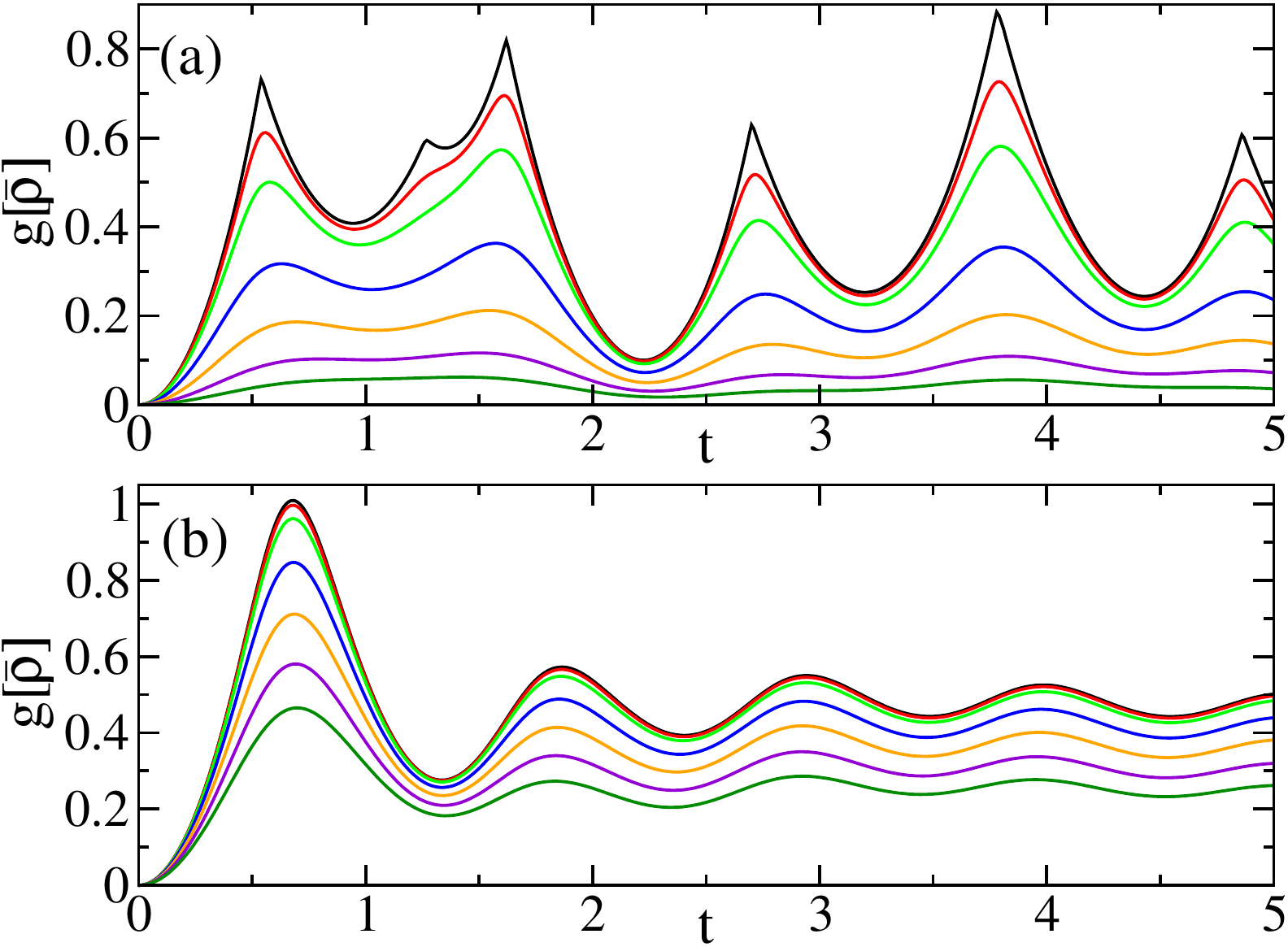}
    \caption{$g[\bar{\rho}]$ for a ramp from $h_i=-2$ to $h_f=2$ for a system of length $L=500$ with $\gamma=1$, $\xi=0,0.05,0.1,0.2,0.3,0.4,0.5$ (from top to bottom), and a ramp velocity (a) $v=1$ and (b) $v=5$. In the latter case $v>v_c\approx \pi\gamma^2/\ln 2$ and no DQPTs are present even without noise.}
    \label{Av_DM_2CP}
\end{figure}
In Fig.~\ref{Av_DM_2CP}, the results for the return rate $g[\bar\rho]$ based on the noise-averaged density matrix are shown. In panel (a), the velocity is below the critical velocity and DQPTs are clearly visible in the noiseless case $\xi=0$. At each critical time, the derivative $\dot g[\bar\rho](t)$ jumps. We also see that these DQPTs immediately get smoothed out once even a small amount of noise is added to the ramp. In panel (b), we have $v>v_c$ and there are no DQPTs even in the clean, noiseless case. In both cases $g[\bar\rho]$ decreases monotonically with increasing noise level and will eventually converge to $g[\bar\rho]\equiv 0$ for $\xi\to\infty$ as expected based on the general discussion in the previous section. Comparing the two ramps it is also clear that the same level of noise has less of an effect on the faster ramp.
\begin{figure}[htp!]
    \includegraphics[width=0.99\columnwidth]{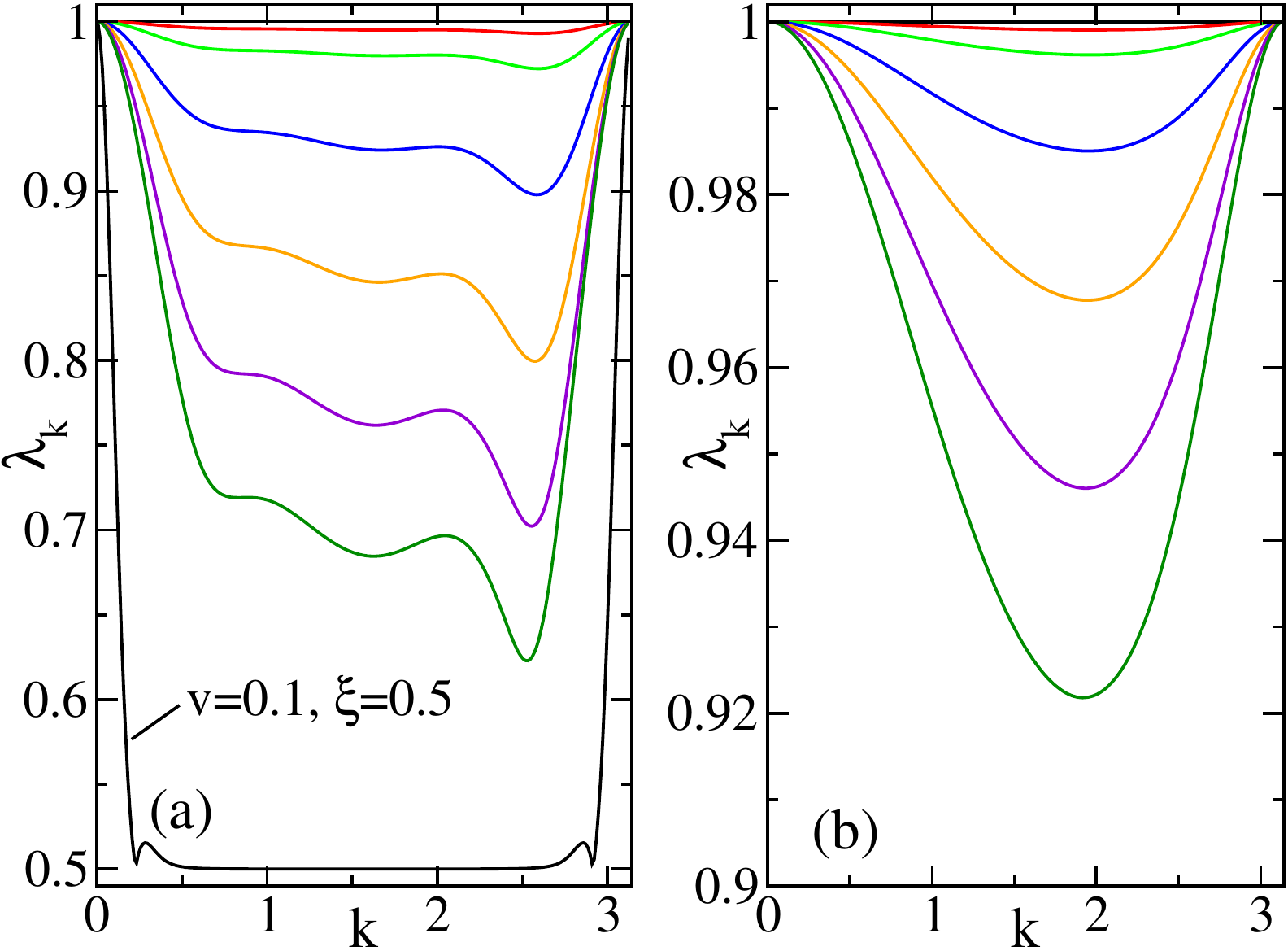}
    \caption{Corresponding largest eigenvalues $\lambda_k$ of $\bar{\rho}^f(0)$ for the return rates $g[\bar{\rho}]$ shown in Fig.~\ref{Av_DM_2CP}. In panel (a), $\lambda_k$ for a slower ramp with $v=0.1$ and $\xi=0.5$ is shown in addition, demonstrating that for $\xi\lesssim\gamma$ and $\xi^2/v\gg1$ the density matrix $\bar\rho$ is driven towards the completely mixed state for all $k$ with $\gamma_k\neq 0$.}
    \label{Av_DM_2CP_EV}
\end{figure}
This can also be clearly seen in Fig.~\ref{Av_DM_2CP_EV} where the largest eigenvalue $\lambda_k$ of $\bar\rho_k(0)$ is shown for different noise levels. With increasing noise, the noise-averaged density matrix  for the slow ramp shown in panel (a) is driven towards the completely mixed state with $\lambda_k=1/2$ for all $k$ with $\gamma_k\neq 0$. On the other hand, the noise-averaged density matrix $\bar\rho_k(0)$ is still quite close to a pure state for the fast ramp and the noise levels considered in Fig.~\ref{Av_DM_2CP_EV}(b). Here, the noise simply has much less time to act and drive the system towards a mixed state. This is consistent with the general qualitative criteria for reaching the mixed state which we have derived in the previous section. 

\subsubsection{Pure state average}
\begin{figure}[htp!]
    \includegraphics[width=0.99\columnwidth]{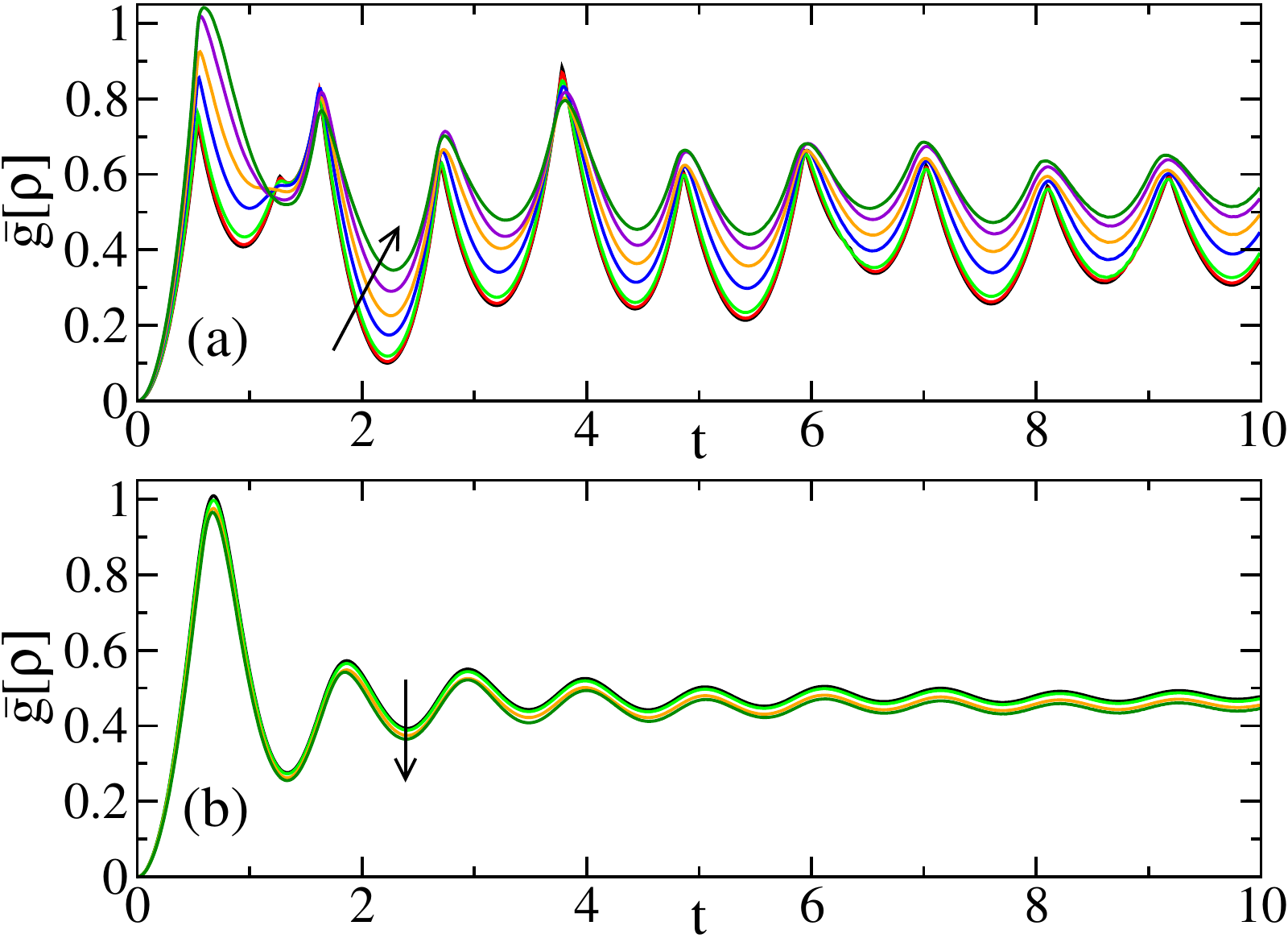}
    \caption{$\bar{g}[\rho]$ for a ramp from $h_i=-2$ to $h_f=2$ for a system of length $L=500$ with $\gamma=1$, $\xi=0,(0.05),0.1,(0.2),0.3,(0.4),0.5$ (in arrow direction), and a ramp velocity (a) $v=1$ and (b) $v=5$. In panel (b), only every second $\xi$ value is plotted. Shown are averages over $200$ samples.}
    \label{Pure_state_2CP}
\end{figure}

Calculating the pure state return rate $g[\rho]$ for each individual noise realization first and then taking the average of these return rates, leads to very different results which are shown in Fig.~\ref{Pure_state_2CP}.
First of all, it is obvious that the same noise level affects $\bar g[\rho]$ much less than $g[\bar\rho]$. Also, for the slower ramp shown in Fig.~\ref{Pure_state_2CP}(a), relatively sharp features do remain for small noise levels and the positions of most of the DQPTs in the clean case can still be inferred from the noise-averaged $\bar g[\rho]$ even for fairly large noise levels. We also note that in contrast to $g[\bar\rho]$ the changes as a function of the noise level are non-monotonic. 
\begin{figure}[htp!]
    \includegraphics[width=0.99\columnwidth]{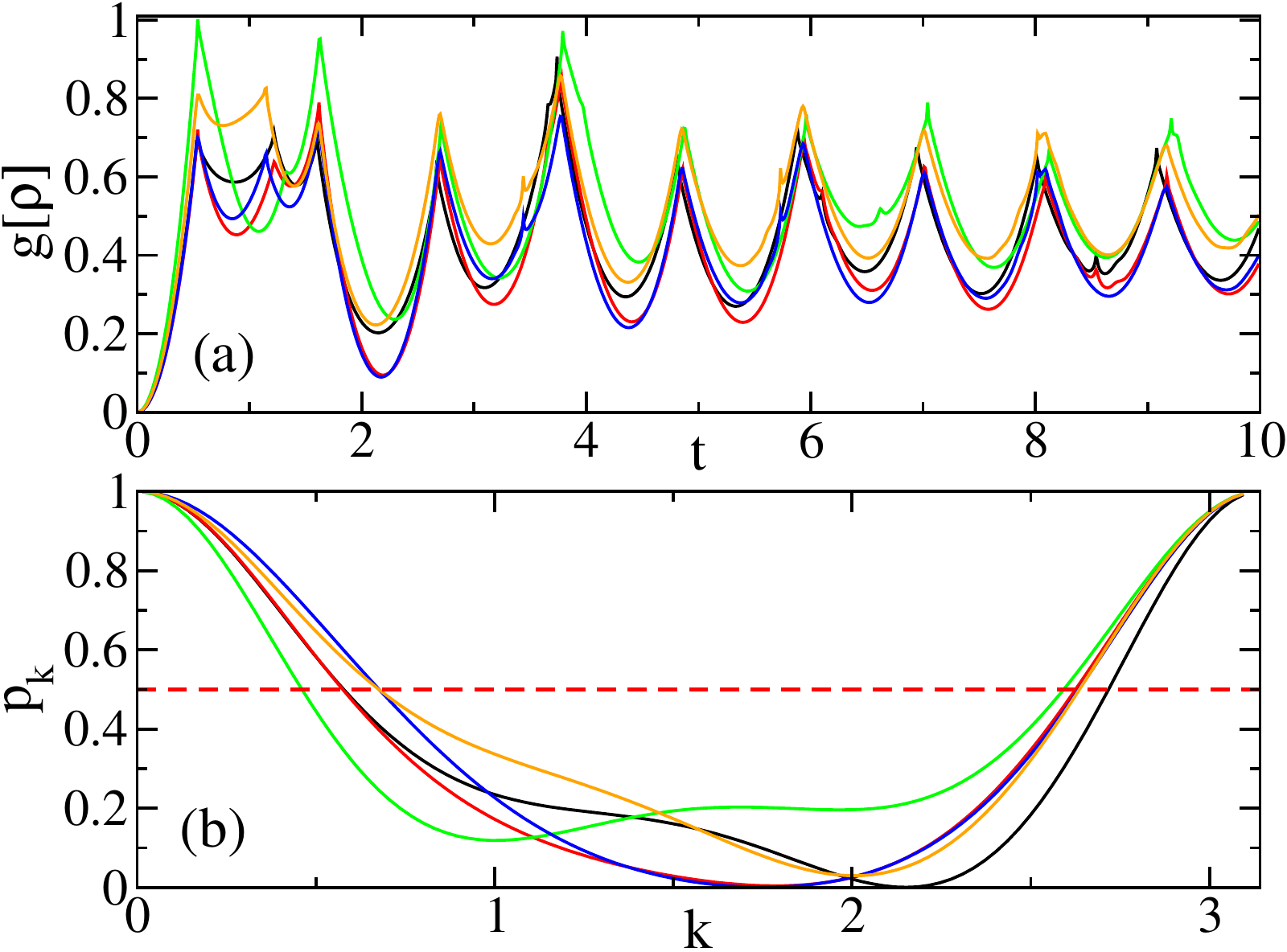}
    \caption{(a) $g[\rho]$ for a ramp from $h_i=-2$ to $h_f=2$ for a system of length $L=500$ with $\gamma=1$, $\xi=0.2$, and $v=1$. 5 different noise realizations are shown. (b) The corresponding $p_k$ always crosses $1/2$ twice at some critical $k_{1,2}^*$ leading to DQPTs even in the presence of noise.}
    \label{Ind_samples_2CP}
\end{figure}
To better understand why the positions of the DQPTs can be inferred from $\bar g[\rho]$, we show in Fig.~\ref{Ind_samples_2CP}(a) individual noise realizations for the slow quench with $v=1$. All realizations do show DQPTs and we find this to be true for almost every realization independent of the strength of the noise. I.e., noise typically does not destroy DQPTs if one considers individual noise realizations. This is important from an experimental point of view and shows that sharp DQPTs are observable even in imperfect systems. The reason why this is the case can be understood from Eq.~\eqref{eq:DFE} which shows that DQPTs occur if there are momenta $k^*$ with $p_{k^*}=1/2$. The $p_k$ curves for the individual noise realizations shown in Fig.~\ref{Ind_samples_2CP}(b) demonstrate that while noise does alter the $p_k$-curves, they all cross $1/2$ at two critical momenta $k_{1,2}^*$ leading to critical times which are determined by Eq.~\eqref{eq:criticaltimes}. Because the momenta $k_{1,2}^*$ are slightly shifted between different noise realizations, the critical times $t^*$ also vary from sample to sample. It is this difference in the critical times which leads to the disappearance of sharp DQPTs when considering the noise-averaged return rate, see Fig.~\ref{Ind_samples_2CP}(a). However, since the variations in $k_{1,2}^*$ remain small even for moderate noise levels, $\bar g[\rho]$ still has distinct peaks which allow to infer the positions of almost all the DQPTs in the clean limit. We also note that the mixing rate is given by $\gamma_{k=0,\pi}=0$ for the XY model and that for a ramp across both critical points we always have $p_{k=0,\pi}=1$. It is therefore in principle possible to have individual noise realizations where $p_k\neq 1/2$ for all momenta $k$. Such a realization thus would not show any DQPTs. Empirically, we find however that even for strong noise this almost never happens.

\subsection{Noisy ramp across a single critical point}
In addition to the ramps across both critical points considered in the section above, we also want to consider ramps across only one of the critical points. An important difference is that in the latter case $p_{k=0}=0$ and $p_{k=\pi}=1$ or vice versa. This guarantees that there will be at least one momentum $k^*$ with $p_{k^*}=1/2$. Therefore every single pure state return rate $g[\rho]$ will show DQPTs no matter what the speed of the ramp or the strength of the noise is. 

\subsubsection{Averaged density matrix}
We start by first considering again the return rate based on the noise-averaged density matrix. 
\begin{figure}[htp!]
    \includegraphics[width=0.99\columnwidth]{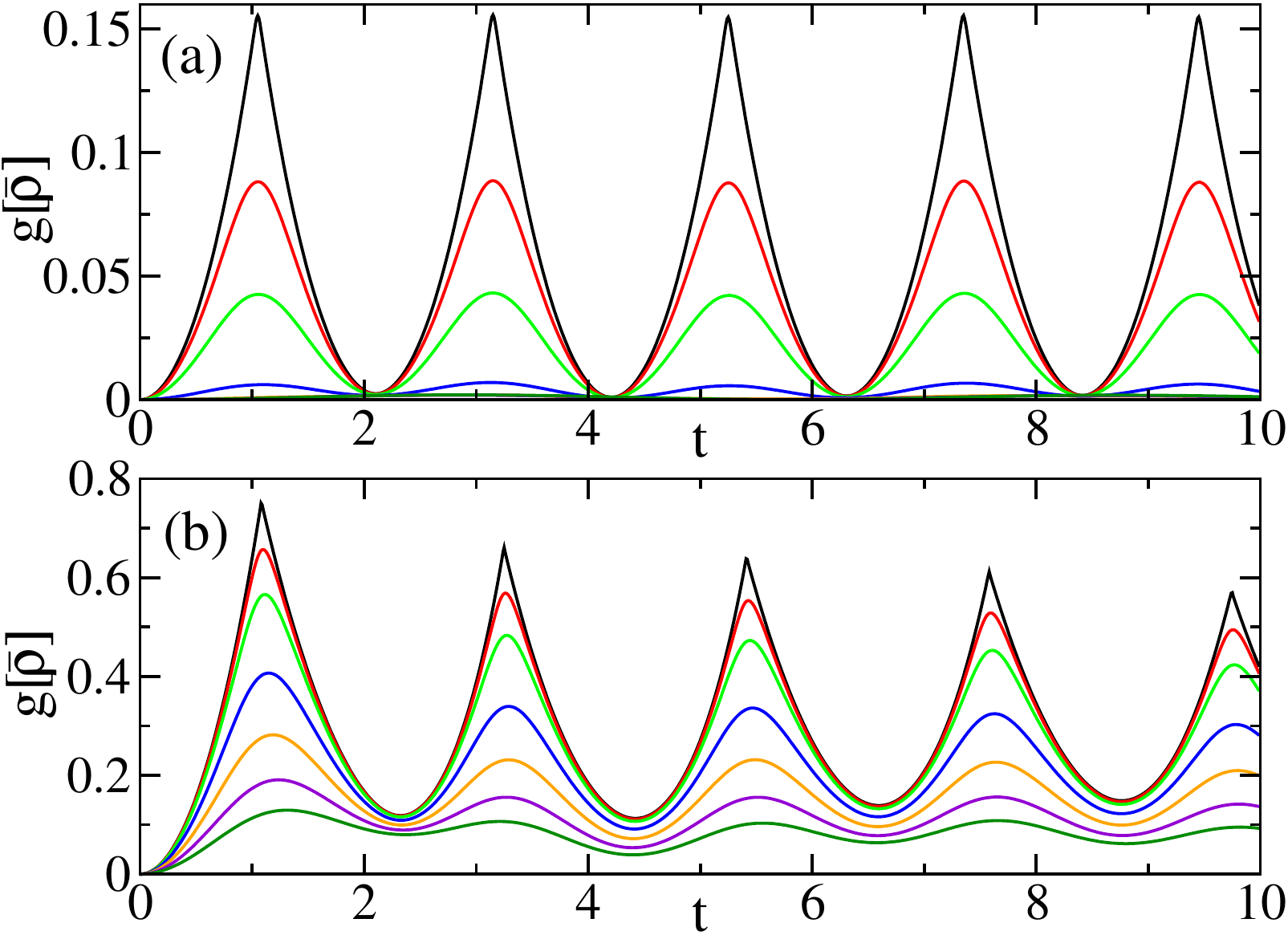}
    \caption{$g[\bar{\rho}]$ for a ramp from $h_i=-2$ to $h_f=0.5$ for a system of length $L=500$ with $\gamma=1$, $\xi=0,0.05,0.1,0.2,0.3,0.4,0.5$ (from top to bottom), and a ramp velocity (a) $v=0.1$ and (b) $v=1$.}
    \label{Av_DM_SCP}
\end{figure}
As shown in Fig.~\ref{Av_DM_SCP}, there are now DQPTs both for the slow and for the fast quench in the noiseless case. Increasing the noise level leads to a monotonic decrease in $g[\bar\rho](t)$. At the same noise level, the effect of the noise is greater for the slower ramp.

\subsubsection{Pure state average}
\begin{figure}[htp!]
    \includegraphics[width=0.99\columnwidth]{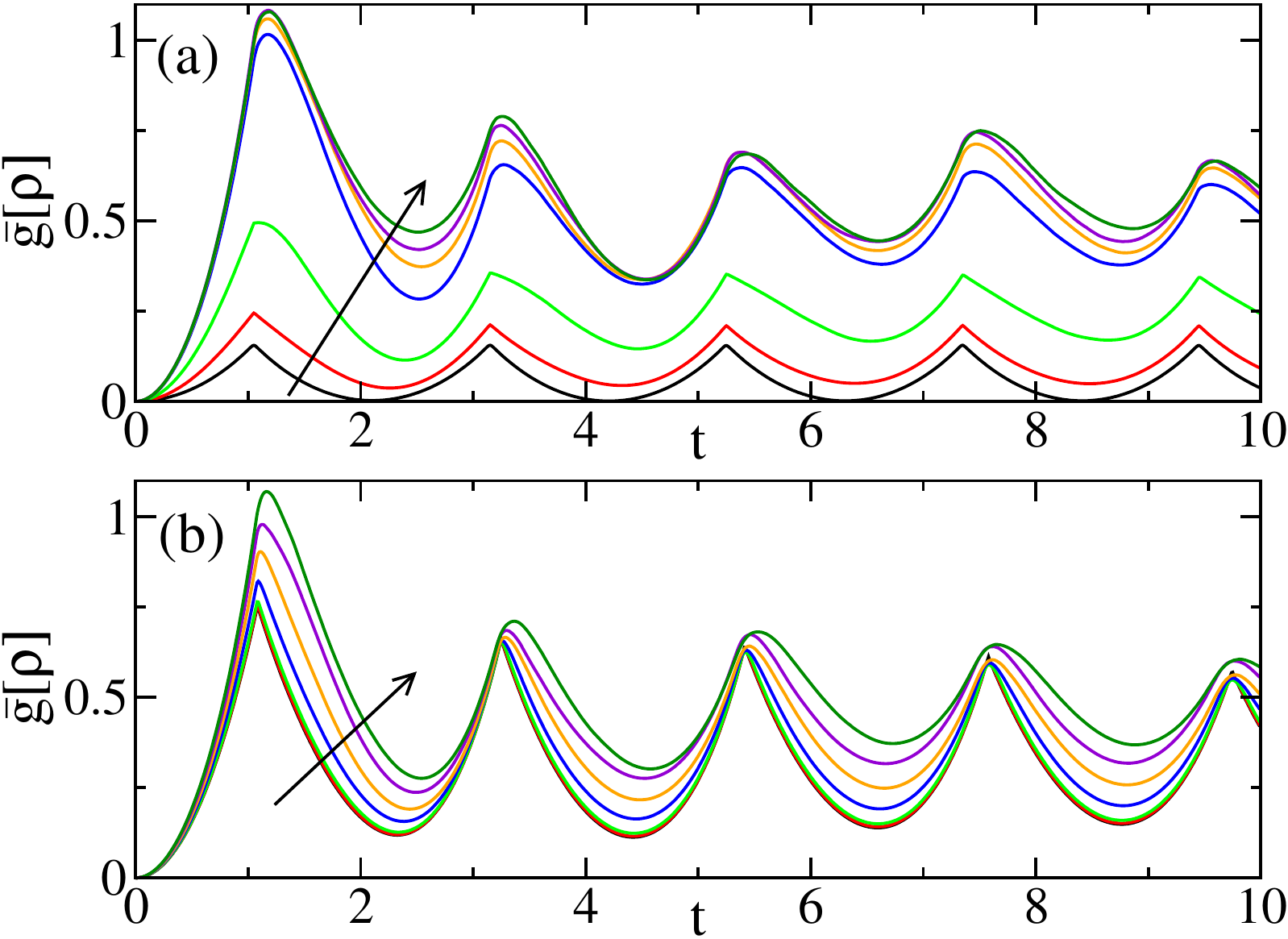}
    \caption{$\bar{g}[\rho]$ for a ramp from $h_i=-2$ to $h_f=0.5$ for a system of length $L=500$ with $\gamma=1$, $\xi=0,0.05,0.1,0.2,0.3,0.4,0.5$ (in arrow direction), and a ramp velocity (a) $v=0.1$ and (b) $v=1$. Overages over $200$ samples have been taken in each case.}
    \label{Pure_state_SCP}
\end{figure}

\begin{figure}[htp!]
    \includegraphics[width=0.99\columnwidth]{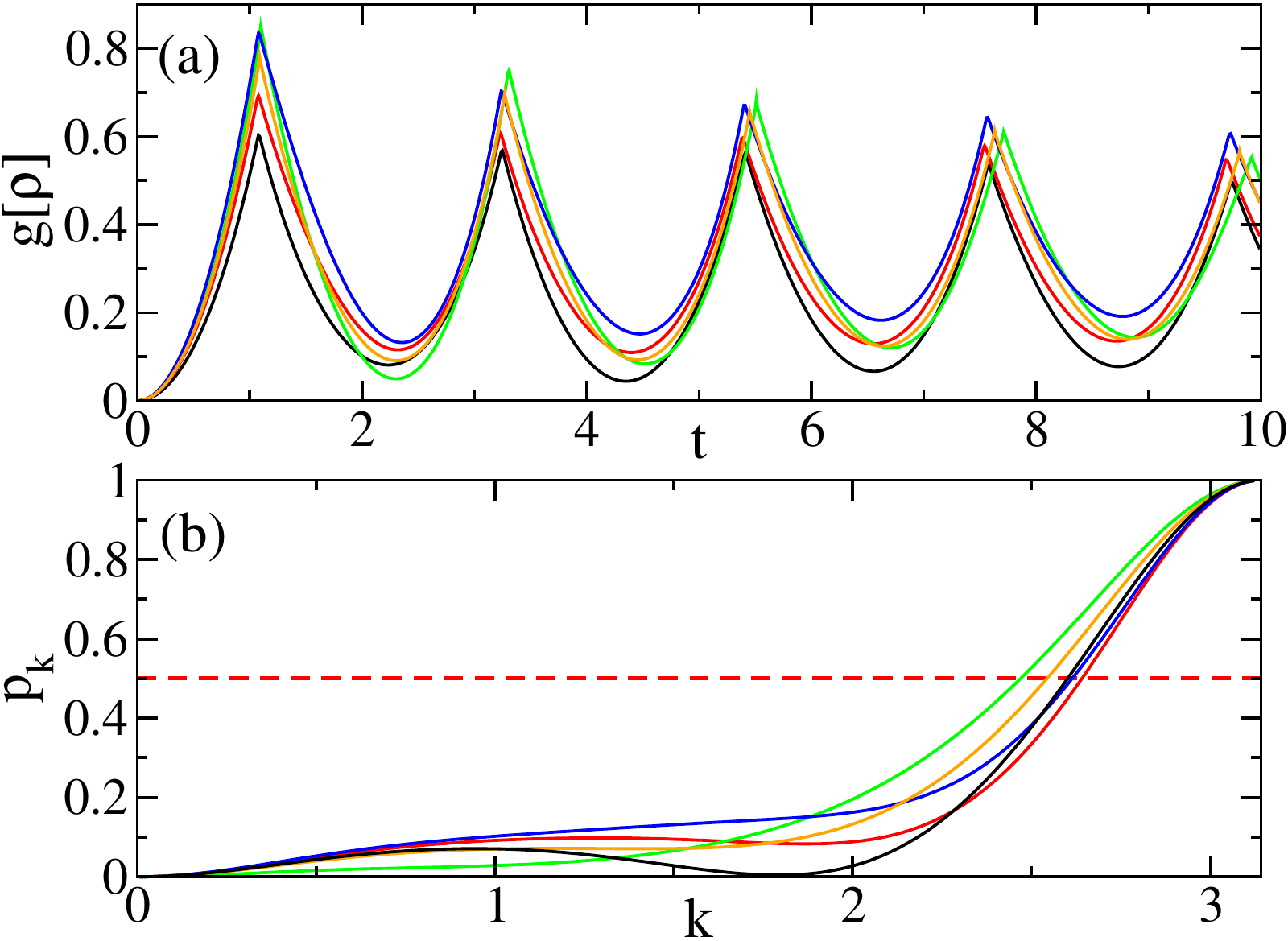}
    \caption{(a) $g[\rho]$ for a ramp from $h_i=-2$ to $h_f=0.5$ for a system of length $L=500$ with $\gamma=1$, $\xi=0.2$, and $v=1$. 5 different noise realizations are shown. (b) The corresponding $p_k$ always crosses $1/2$ at some critical $k^*$ leading to DQPTs even in the presence of noise.}
    \label{Ind_samples_SCP}
\end{figure}

The corresponding results for the noise-averaged return rate $\bar g[\rho]$ are shown in Fig.~\ref{Pure_state_SCP}.
As for the ramp across both critical points, we find again that relatively sharp features remain present for small and intermediate noise levels and that the positions of the DQPTs in the clean case can still be identified even for strong noise. The reason becomes obvious when looking at the individual noise realizations shown in Fig.~\ref{Ind_samples_SCP}.
As shown in panel (b), we have $p_{k=0}=0$ and $p_{k=\pi}=1$ so that for each realization there is a $k^*$ with $p_{k^*}=1/2$. Also, for the moderate noise level considered in this figure, $k^*$ only shifts slightly between different realizations, leading to criticial times $t^*$ which are similar, see panel (a). We conclude that a ramp across a single critical point in the XY model is an ideal scenario to observe DQPTs experimentally because in this case they are guaranteed to occur no matter how noisy the ramp is. 

\subsection{Noise in the energy levels of the final Hamiltonian}
As a second aspect of the influence of noise on the non-equilibrium dynamics of quantum systems, we study fluctuations in the energy levels of the post-ramp Hamiltonian. 
One approach to study the effects of the environment on a quantum system \cite{Joos2013,Zurek2003} is through stochastic fluctuations in a system's observable, 
which is described by the Kubo-Anderson spectral diffusion process \cite{Anderson1954,Kubo1954,Kubobook,Masashi2010}. 
To investigate the impact of energy level fluctuations on DQPTs, we study a general two-band model with Hamiltonian \eqref{eq:Nambu} with a post-ramp Hamiltonian $H^f_{k}(t)=-\epsilon^{f}_{k}\sigma^{z}+r(t)\sigma^{z}$ where $r(t)$ represents white noise \cite{Anderson1954,Kubo1954,Kubobook}. We will consider again both the return rate based on the noise-averaged density matrix $g[\bar\rho]$ as well as the return rate $\bar g[\rho]$ which is averaged over individual noise realizations.

\subsubsection{Noise-averaged density matrix}
For this type of noise, the theoretical advantage of working with the noise-averaged density matrix becomes particularly obvious. Regardless of whether the ramp crosses a single critical point or two critical points, 
the noise master equation, Eq.~(\ref{eq:master}),  
is exactly solvable and yields a closed-form expression for the return rate (see App.~\ref{App_B})
%
\begin{equation}
\bl
g[\bar\rho]=-\frac{1}{\pi} \int_{0}^{\pi}\ln
\left[
1-4p_{k}(1-p_{k})\Big(\frac{1}{2}-\frac{F(t)}{2}
+F(t)\sin^{2}(\epsilon_{k}^{f}t)\Big)
\right] \ dk,
\el
\label{eq:DFEN}
\end{equation}	    	
%
with decoherence factor $F(t)=\exp(-2\xi^2t)$. It is immediately obvious that the condition $\frac{1}{2}-\frac{F(t)}{2}+F(t)\sin^{2}(\epsilon_{k}^{f}t)=1$ cannot be fulfilled except in the case without noise, $\xi=0$, in which case Eq.~\eqref{eq:DFEN} reduces to Eq.~\eqref{eq:DFE}. Stochastic fluctuations in the energy levels 
of the post-ramp Hamiltonian always prevent the occurrence of DQPTs. We note that even if the environmental noise $r(t)$ follows an Ornstein-Uhlenbeck process (correlated noise), the return rate can still be expressed in closed form with a decoherence factor which can be determined exactly, see App.~\ref{App_B}. 
\begin{figure}[htp!]
    \includegraphics[width=0.99\columnwidth]{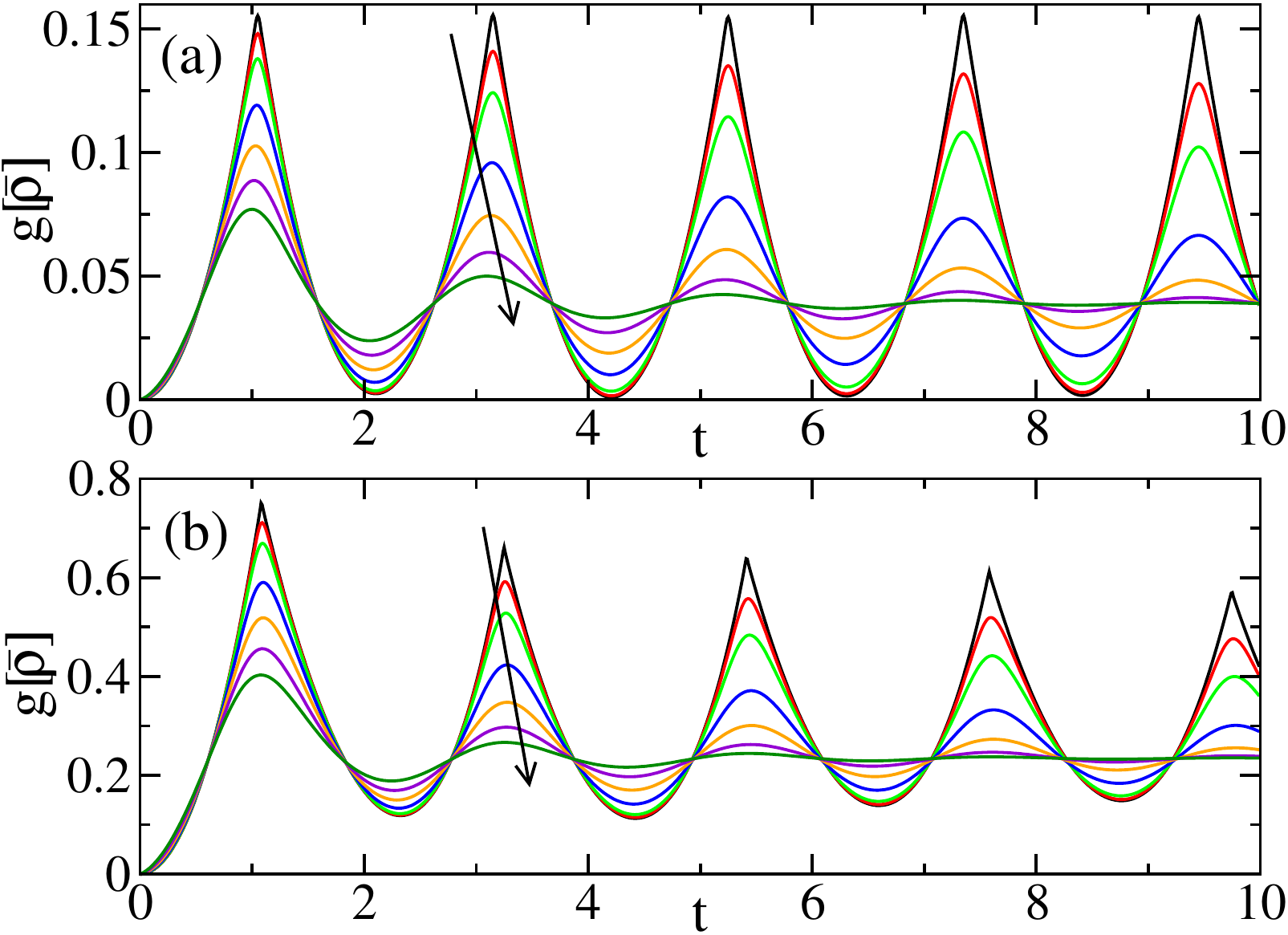}
    \caption{$g[\bar{\rho}]$ for a noiseless ramp from $h_i=-2$ to $h_f=0.5$ for a system of length $L=500$ with $\gamma=1$ followed by a time evolution with noise in the energies of strength $\xi=0,0.05,0.1,0.2,0.3,0.4,0.5$ (in arrow direction) for a ramp velocity (a) $v=0.1$ and (b) $v=1$.}
    \label{Av_DM_SCP_energy}
\end{figure}
Examples for $g[\bar\rho]$ when fluctuations in the energy levels of the post-ramp Hamiltonian are present are shown in Fig.~\ref{Av_DM_SCP_energy}. In contrast to the case with noise during the ramp, $g[\bar\rho]$ is no longer a monotonically decreasing function of the noise level for all times. Rather, the amplitude of oscillations is monotonically decreasing and there are universal crossing points in time where $g[\bar\rho]$ is independent of the noise $\xi$. From Eq.~\eqref{eq:DFEN} one can see that for $\xi\to\infty$ the integrand becomes $\ln[1-2p_k(1-p_k)]$, thus only depends on the state after the noiseless ramp, and will in general result in $g[\bar\rho]=\mbox{const}\neq 0$.

\subsubsection{Pure state average}
For the case of noise in the energy levels of the final Hamiltonian, we expect the results for $g[\bar\rho]$ based on calculating the noise-averaged density matrix and the results for $\bar g[\rho]$ based on averaging over the pure-state return rates to be qualitatively similar. This is in contrast to the noisy ramp discussed in the previous section where they are qualitatively different. The reason is that now in both cases the state after the noiseless ramp is the same pure state described by $p_k$ and that Eq.~\eqref{eq:DFEN} reduces to the pure state formula \eqref{eq:DFE} in the case without noise. 
\begin{figure}[htp!]
    \includegraphics[width=0.99\columnwidth]{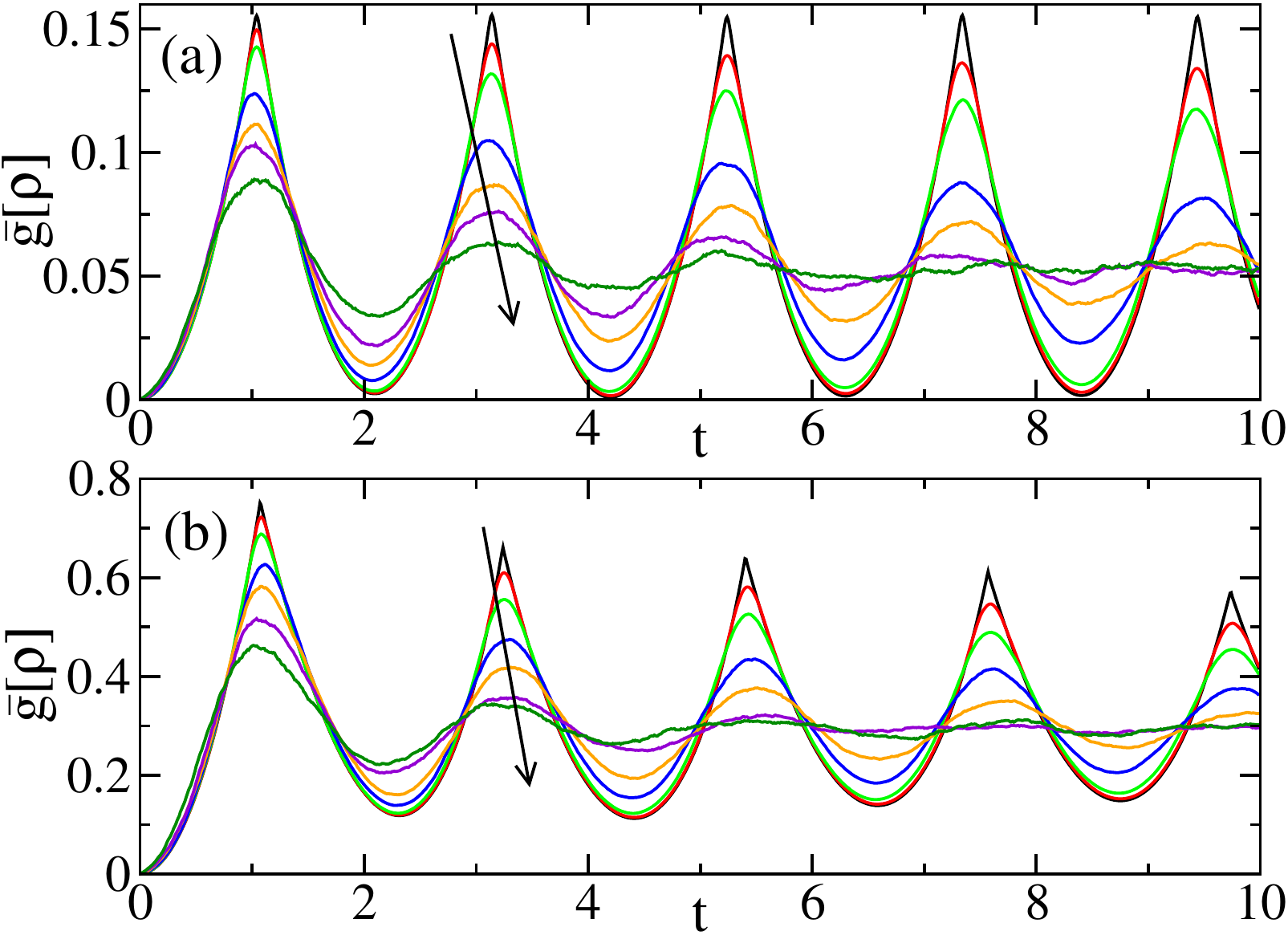}
    \caption{$\bar{g}[\rho]$ for a noiseless ramp from $h_i=-2$ to $h_f=0.5$ for a system of length $L=500$ with $\gamma=1$ followed by a time evolution with noise in the energy levels of strength $\xi=0,0.05,0.1,0.2,0.3,0.4,0.5$ (in arrow direction) and a ramp velocity (a) $v=0.1$ and (b) $v=1$. Overages over $400$ samples have been taken in each case.}
    \label{Pure_state_SCP_energy}
\end{figure}
A comparison of $\bar g[\rho]$ shown in Fig.~\ref{Pure_state_SCP_energy} with $g[\bar\rho]$ shown in Fig.~\ref{Av_DM_SCP_energy} confirms this expectation.

\section{Summary and conclusions}
\label{Sec_Concl}
We have studied the effects of noise on quantum ramps in two-band fermionic models. More specifically, we have investigated the Loschmidt echo and the corresponding return rate as a function of time $t$ after the end of the ramp. There are two principal ways a noise average can be performed: (i) One can calculate the noise-averaged density matrix $\bar\rho$ which will be a mixed state in general and then use a generalized Loschmidt echo $g[\bar\rho]$ for density matrices. (ii) One calculates the return rate for each noise realization individually and averages the results to obtain $\bar g[\rho]$. The advantage of the former method is that a master equation of Lindblad type can be obtained for $\bar\rho$ which, as we have demonstrated, can be solved analytically in some cases and efficiently numerically in others. The results show that noise in this formalism always destroys DQPTs although for small noise levels DQPTs are merely smoothed out and can still be identified. The second method might be more relevant experimentally because it only requires the measurement of $|\langle\Psi^f(0)|\Psi^f(t)\rangle|$. Nevertheless, even this might not always be straightforward because the state $|\Psi^f(0)\rangle$ after the ramp is in general not a simple product state. Interestingly, we find that for two-band models with mixing rate $\gamma_{k_1,k_2}=0$ and $p_{k_1}=0(1),\; p_{k_2}=1(0)$ the existence of at least one critical momentum $k^*$ with $p_{k^*}=1/2$ is guaranteed. In this case DQPTs for individual noise realizations always occur no matter how strong the noise is. However, the corresponding critical times $t^*$ are different for each realization so that DQPTs in $\bar g[\rho]$ are also always smoothed out. However, they do remain quite sharp for small to intermediate noise levels implying that DQPTs in this situation are robust and can be studied experimentally in imperfect systems.

\acknowledgments
J.S.~acknowledges support by NSERC via the Discovery grants program.

\appendix
\section{A noiseless ramp}
\label{App_A}
Here we briefly review the case of a noiseless ramp in the XY chain. Consider the Hamiltonian $H^{0}_k({t})$ which governs the fermionic modes of the Jordan-Wigner transformed quantum $XY$ chain, Eq.~\eqref{eq:Nambu}, with $h^{z}_k(t) = h_0(t)-\cos(k)$ and $\gamma_k = \gamma\sin(k)$. During a ramp in the time interval $[t_i,t_f)$, the transverse magnetic field $h_{0}(t)$ changes from an initial value $h_0(t_i)=h_i$ to a final value $h_0(t_f)=h_f$ such that $h_0({t}) = h_f + vt$, and $v>0$. Rewriting $H^{0}_k(t)$ in the form of a Landau-Zener model \cite{Landau,Zener}, one obtains $H^{0}_k({t}) = v\tau_k \sigma_z + \gamma_k \sigma_x$, with $\tau_k = h^z_k(t) /v$ a mode-dependent time variable. The probability that the $k$-th mode is found in the excited state of the final Hamiltonian at the end of the ramp is then given by \cite{Vitanov1999,TorosovVitanov}
%
\bea
\label{eq:SMA1}
p_k=  \left|U_{11}\sin\Big(\frac{\Theta(\tau_{k,f})}{2}\Big)+ U_{21}\cos\Big(\frac{\Theta(\tau_{k,f})}{2}\Big)\right|^2,
\eea
%
where,
%
\begin{widetext}
{\small
\bea
\no
U_{11}&=&\frac{\Gamma(1-i\frac{\Delta^2}{2v})}{\sqrt{2\pi}}
\Big[D_{i\frac{\Delta^2}{2v}-1}\Big(\sqrt{2v}e^{-i\pi/4}\tau_{k,i}\Big)D_{i\frac{\Delta^2}{2v}}\Big(-\sqrt{2v}e^{-i\pi/4}\tau_{k,f}\Big)
+D_{i\frac{\Delta^2}{2v}-1}\Big(-\sqrt{2v}e^{-i\pi/4}\tau_{k,i}\Big)D_{i\frac{\Delta^2}{2v}}\Big(\sqrt{2v}e^{-i\pi/4}\tau_{k,f}\Big)\Big],\\ \no
U_{21}&=&\frac{\Delta\Gamma(1-i\frac{\Delta^2}{2v})}{\sqrt{\pi v}}e^{-i\pi/4}
\Big[D_{i\frac{\Delta^2}{2v}-1}\Big(\sqrt{2v}e^{-i\pi/4}\tau_{k,i}\Big)D_{i\frac{\Delta^2}{2v}-1}\Big(-\sqrt{2v}e^{-i\pi/4}\tau_{k,f}\Big)\\ \no
&&\qquad\qquad\qquad\qquad\quad -D_{i\frac{\Delta^2}{2v}-1}\Big(-\sqrt{2v}e^{-i\pi/4}\tau_{k,i}\Big)D_{i\frac{\Delta^2}{2v}-1}\Big(\sqrt{2v}e^{-i\pi/4}\tau_{k,f}\Big)\Big],
\eea
}
\end{widetext}
%
with $\tan(\Theta(\tau_k))=\gamma_k/(v\tau_k)$ and $\Theta\in[0,2\pi]$, $\tau_{k,i}=(\cos(k)-h_i)/v$, $\tau_{k,f}=(\cos(k)-h_f)/v$, $\Delta=\gamma_k$, $D_{\zeta}(z)$ the parabolic cylinder function \cite{szego1954,abramowitz1988}, and $\Gamma(x)$ the Euler Gamma function.
For a ramp from $h_i\to -\infty$ to $h_f\to\infty$ this formula reduces to the well-known Landau-Zener transition probability $p_k=\exp(-\pi(\gamma\sin(k))^2/v)$. DQPTs in this limit thus only occur if $v<v_c=\pi \gamma^2/\ln 2.$

For a ramp from $h_i\to -\infty$ across the critical field $h=-1$ to some final value $-1<h_f<1$ in the ferromagnetic phase, on the other hand, the transition probability $p_k=p_k(h_i,h_f)$ 
for modes $k\sim 0$ will be small, $p_k\ll 1/2$, while it will be given by $p_k\lesssim 1$ for modes close to the gap-closing point $k\sim\pi$ \cite{Zamani2024,Sharma2016,Jafari2024}. Given these two limiting cases, the continuity of $p_k$ as a function of $k$ in the thermodynamic limit implies that there exists at least one critical mode $k^{\ast}$ with equal amplitudes $p_{k^{\ast}}=1/2$ for the occupation of the lower and upper levels. This is the
mode that triggers the appearance of DQPTs at critical times. In other words, for the XY model DQPTs are always present for a ramp 
across a single critical point, even in the limit of a sudden quench \cite{Zamani2024,Sharma2016,Jafari2024}. 

\section{Energy level fluctuations in the post-ramp Hamiltonian}
\label{App_B}
Random unitary dynamics emerges in quantum mechanics as an effective way for characterizing the evolution of 
systems that interact with their environments or external fields. The original idea was proposed by Caldeira and Leggett 
to examine the effective dynamics of collections of spins interacting with bosonic baths \cite{CALDEIRA1983}.
One of the simplest methods that may serve as a paradigm for the impact of an environment on the quantum system is the Kubo-Anderson spectral diffusion process \cite{Anderson1954,Kubo1954,Kubobook,Masashi2010} where the effect of the environment on the quantum system is described by stochastic fluctuations in a system's observable.
In this context, we assume that the post-ramp energy levels show stochastic fluctuations. Therefore, the post-ramp Hamiltonian in the diagonal basis can be written as $H^f_{k}(t)=-\epsilon^{f}_{k}\sigma^{z}+r(t)\sigma^{z}$ where $r(t)$ represents noise processes and the density matrix $\rho_{k}(t)$ for a mode $k$ at $t=0$ takes the form 
%
\begin{equation}
\bl
\label{eq:DMI}
\rho_k(t=t_f=0) =| \psi_{k}(h_f) \rangle \langle \psi_{k}(h_f)|=
\left[
\begin{array}{cc}
|v_k|^2 & v_k u^{\ast}_k \\
v^{\ast}_k u_k & |u_k|^2\\
\end{array}
\right]
\el
\end{equation}
%
with $|v_k|^2+|u_k|^2=1$. It is straightforward to show that in the colored noise process with $\langle r({t})r({t}')\rangle=(\xi^{2}/2\tau_n)\exp(-|{t}-{t}'|/\tau_n)$ 
the dynamical evolution of the density matrix elements $\rho_{k,ij}(t)$ can be written as
%
\bea
\label{eq:DMTCN}
\frac{d}{dt}\rho_{k,11}(t) &=& 0, \\
\no
\frac{d}{dt}\rho_{k,12}(t) &=& 2 i\epsilon^{f}_{k}\rho_{k,12}(t)-\frac{2\xi^2}{\tau_n}\int_{0}^{t} e^{-(t-t')/\tau_n}\rho_{k,12}(t') dt', \\
\no
\frac{d}{dt}\rho_{k,21}(t) &=& -2 i\epsilon^{f}_{k}\rho_{k,21}(t)-\frac{2\xi^2}{\tau_n}\int_{0}^{t} e^{-(t-t')/\tau_n}\rho_{k,21}(t') dt',\\
\no
\frac{d}{dt}\rho_{k,22}(t) &=& 0, 
\eea
%
which reduces to 
%
\bea
\label{eq:DMTWN}
\frac{d}{dt}\rho_{k,11}(t) &=& 0, \\
\no
\frac{d}{dt}\rho_{k,12}(t) &=& 2 (i\epsilon^{f}_{k}-\xi^2)\rho_{k,12}(t), \\
\no
\frac{d}{dt}\rho_{k,21}(t) &=& -2 (i\epsilon^{f}_{k}+\xi^2)\rho_{k,21}(t),\\
\no
\frac{d}{dt}\rho_{k,22}(t) &=& 0, 
\eea
%
for white noise $\langle r({t})r({t}')\rangle = \xi^2\delta(t-t')$. By using the Laplace transform of Eq.~(\ref{eq:DMTCN}), and its inverse the density matrix elements can be expressed as
%
\bea
\no
\rho_{k,11}(t) &=& \rho_{k,11}(0), \\
\rho_{k,12}(t) &=& \frac{e^{(i\epsilon^{f}_{k}-\lambda/2)t}}{\delta}\\ \no
&\times&\left[\delta\cosh(\delta)+(2i\epsilon^{f}_{k}+\lambda)\sinh(\delta)\right]\rho_{k,12}(0), \\
\no
\rho_{k,21}(t) &=& \frac{e^{-(i\epsilon^{f}_{k}+\lambda/2)t}}{\delta^{\ast}}\\ \no
&\times&\left[\delta^{\ast}\cosh(\delta^{\ast})-(2i\epsilon^{f}_{k}-\lambda)\sinh(\delta^{\ast})\right]\rho_{k,21}(0),\\
\no
\rho_{k,22}(t) &=& \rho_{k,22}(0), 
\eea
%
with $\lambda=\tau_n^{-1}$ and $\delta=\sqrt{(2i\epsilon^{f}_{k}+\lambda)^2-8\xi^2\lambda}$. 
For white noise the above equations are simplified to 
%
\bea
\no
&& \rho_{k,11}(t) = \rho_{k,11}(0),~~ \rho_{k,12}(t) = e^{2(i\epsilon^{f}_{k}-\xi^2)t} \rho_{k,12}(0),\\ \no
&& \rho_{k,21}(t) = e^{-2(i\epsilon^{f}_{k}+\xi^2)t} \rho_{k,21}(0),~~\rho_{k,22}(t) = \rho_{k,22}(0). 
\eea
%
The dynamical evolution of the density matrix $\rho(t)$ can be written as
%
\bea
\bl
\label{eq:DMF}
\rho_k(t) = 
\left[
\begin{array}{cc}
|v_k|^2 & v_k u^{\ast}_k F(t) e^{2i\epsilon^{f}_{k}}\\
v^{\ast}_k u_k F^{\ast}(t) e^{-2i\epsilon^{f}_{k}}& |u_k|^2\\
\end{array}
\right],
\el
\eea
%
where 
$F(t)=\exp[-(i\epsilon^{f}_{k}+\lambda/2)t)][\delta\cosh(\delta)+(2i\epsilon^{f}_{k}+\lambda)\sinh(\delta)]/\delta$
and
$F(t)=\exp(-2\xi^2t)$
are the decoherence factors for colored and white noise, respectively.

Finally, the Loschmidt echo for a $2\times 2$ density matrix $\rho(t)$ for the case that $\rho(0)$ is a pure state can be written as $|{\cal{L}}_{k}|^{2}=Tr(\rho(0)\rho(t))+2\Big(\det(\rho(t))\det(\rho(0))\Big)^{1/2}$ \cite{Sedlmayr2018}.
This quantity measures the degree of distinguishability between the two quantum states $\rho(t)$ and $\rho(0)$. Substituting $\rho(t)$ and $\rho(0)$ defined above leads to 
%
{\small
\begin{eqnarray}
&&|{\cal{L}}_{k}|^{2}=1-4p_{k}(1-p_{k}) \\*[0.15cm]
&\times&\Big(\frac{1-\textrm{Re}(F(t))}{2}+\frac{\textrm{Im}(F(t))}{2}\sin(2\epsilon_{k}^{f}t) + \textrm{Re}(F(t))\sin^{2}(\epsilon_{k}^{f}t)\Big), \no
\end{eqnarray}
}
%
which reduces to the noiseless case if the decoherence factor is $F(t)=1$. Moreover, when the stochastic fluctuations are uncorrelated (white noise) the decoherence factor is
$F(t)=\exp(-2\xi^2t)$, and therefore $\textrm{Im}(F(t))=0$, which leads to
%
\begin{equation}
|{\cal{L}}_{k}|^{2}=1-4p_{k}(1-p_{k})\Big(\frac{1-F(t)}{2}+ F(t)\sin^{2}(\epsilon_{k}^{f}t)\Big).
\end{equation}\\

%
%


\end{document}